\documentclass[aps,prb,reprint,showpacs,twoside]{revtex4-1}%
\usepackage{amsmath}
\usepackage{fancyhdr}
\usepackage{graphicx}
\usepackage{mathptmx}
\usepackage{subfigure}
\usepackage[colorlinks=true,citecolor=blue,linkcolor=blue,urlcolor=blue,bookmarks=false,breaklinks=true]%
{hyperref}
\usepackage{CJK}
\usepackage{amsfonts}
\usepackage{amssymb}%
\providecommand{\U}[1]{\protect\rule{.1in}{.1in}}
\DeclareMathOperator{\sgn}{sgn}

\fancypagestyle{plain}{
\fancyhf{}
\fancyfoot[C]{\small \thepage}
\fancyhead[C]{\small Published version: \href{\doibase 10.1103/PhysRevB.87.125427}{{Phys. Rev. B}\ \textbf{87},\ {125427} (2013)}}
}

\begin{document}
\begin{CJK*}{Bg5}{bsmi}
\title{Theory of carrier density in multigated doped graphene sheets with quantum correction}
\author{Ming-Hao Liu (¼B©ú»¨)}
\email{minghao.liu.taiwan@gmail.com}
\affiliation{Institut f\"{u}r Theoretische Physik, Universit\"{a}t Regensburg, D-93040
Regensburg, Germany}
\date{\today }
\begin{abstract}
The quantum capacitance model is applied to obtain an exact solution for the
space-resolved carrier density in a multigated doped graphene sheet at zero
temperature, with quantum correction arising from the finite electron
capacity of the graphene itself taken into account. The exact solution is
demonstrated to be equivalent to the self-consistent Poisson-Dirac iteration
method by showing an illustrative example, where multiple gates with
irregular shapes and a nonuniform dopant concentration are considered. The
solution therefore provides a fast and accurate way to compute spatially
varying carrier density, on-site electric potential energy, as well as
quantum capacitance for bulk graphene, allowing for any kind of gating
geometry with any number of gates and any types of intrinsic doping.
\end{abstract}
\maketitle

\pagestyle{fancy}
\fancyfoot[C]{\small \thepage}
\fancyhead[LE]{\small{MING-HAO LIU (¼B©ú»¨)}}
\fancyhead[LO]{\small{THEORY OF CARRIER DENSITY IN MULITIGATED DOPED GRAPHENE \ldots}}
\fancyhead[R]{\small Published version: \href{\doibase 10.1103/PhysRevB.87.125427}{{Phys. Rev. B}\ \textbf{87},\ {125427} (2013)}}

\thispagestyle{plain}

\paragraph*{Introduction.}

Manipulation of carrier density in graphene by electrical gating is one of the
key techniques for graphene electronics. Since the first successful isolation
of monolayer graphene flakes, conductance (resistance) sweep using a single
backgate has been a standard electronic characterization tool for
graphene.\cite{Novoselov2004} Double-gated graphene opens possibilities for
experimental investigations of graphene \emph{pn} and \emph{pnp}
junctions,\cite{Huard2007,Williams2007,Ozyilmaz2007} which allow for
exploration of the interesting physics of Klein paradox \cite{Klein1929} in
graphene.\cite{Katsnelson2006,Stander2009,Young2009} In order to improve the
junction quality, graphene heterojunctions using contactless top gates
\cite{Liu2008c,Gorbachev2008} and embedded local gates \cite{Nam2011,Lee2012}
were proposed and investigated.

More complicated gating geometry is involved in recent proposals for
graphene-based devices, such as a switching device with two
topgates,\cite{Nakaharai2012} graphene transistors with self-aligned gates
made by standard patterning with a regular cross section,\cite{Farmer2010}
core-shell nanowires with round cross sections,\cite{Liao2010} or deposited
films with T-shaped cross sections.\cite{Badmaev2012} Transport through
bilayer graphene with multiple top gates up to eight was recently
investigated;\cite{Miyazaki2012} patterning periodic top gates
\cite{Weiss1989} on graphene to form quasi-one-dimensional superlattice is, in
principle, feasible. Whereas a successful transport simulation relies
decisively on the preciseness of the on-site potential profile, or
equivalently the carrier density profile,\cite{Liu2012a} a more reliable
theory to deal with general gating geometry is, therefore, imperative.

The theory of gate-induced carrier density started from the simplest classical
capacitance model,\cite{Novoselov2004} which regards the
graphene-substrate-backgate as a parallel-plate capacitor and the relevant
carrier density in graphene as the surface charge density (divided by electron
charge $-e$) induced by the gate. Without taking into account the quantum
correction due to the finite capacity of graphene itself for electrons to
reside, this model can be straightforwardly generalized to arbitrary gating
geometry by treating graphene as a perfect conducting plane with fixed zero
potential. A more precise computation of the gate-induced carrier density,
however, needs to take into account the relation between the induced charge
density on graphene and the electric potential energy that those charge
carriers gain, through the graphene density of
states.\cite{Guo2007,Fern'andez-Rossier2007,Fang2007} The solution to the
carrier density with such a correction taken into account requires a
self-consistent iteration process
\cite{Gorbachev2008,Shylau2009,Andrijauskas2012} that may be suitably termed
the Poisson-Dirac method but actually corresponds to the quantum capacitance
model,\cite{Luryi1988} where an exact solution for single-gated pristine
graphene at zero temperature has been derived.\cite{Fang2007}

In this paper, the spatial profile of carrier density in monolayer graphene
due to arbitrary gating and doping is exactly solved within the quantum
capacitance model. The solution has been further tested by comparing with the
self-consistent Poisson-Dirac method, showing very good agreement between the
two and, hence, their equivalence. A numerical example will be illustrated at
the end. Throughout, we will restrict our discussion to bulk graphene at zero
temperature and approximate the energy dispersion within the linear Dirac
model, $E=\pm\hbar v_{F}k,$ which leads to the density of states (per unit
area) linear in energy, $D(E)=2\left\vert E\right\vert /\pi(\hbar v_{F})$. The
carrier density is given by integrating the density of states over the energy,%
\begin{equation}
n(E)=\sgn(E)\frac{1}{\pi}\left(  \frac{E}{\hbar v_{F}}\right)  ^{2},
\label{n(E)}%
\end{equation}
which is the underlying origin of the quantum correction to the gate-induced
graphene carrier density in the following derivations. We are, therefore,
working in the single-particle picture, and the solution within the quantum
capacitance model to be presented is exact in the sense that no iteration is
required during the solution process, as contrary to the following
Poisson-Dirac method.

\paragraph*{Self-consistent Poisson-Dirac iteration method.}

\begin{figure}[t]
\subfigure[]{
\includegraphics[height=3.5cm]{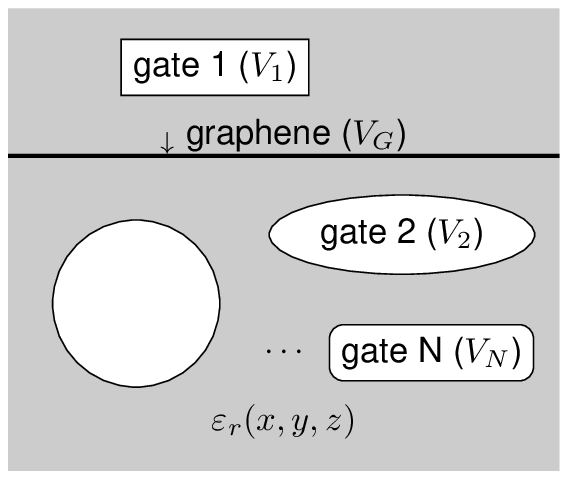}
\label{fig1a}} \subfigure[]{
\includegraphics[height=3.5cm]{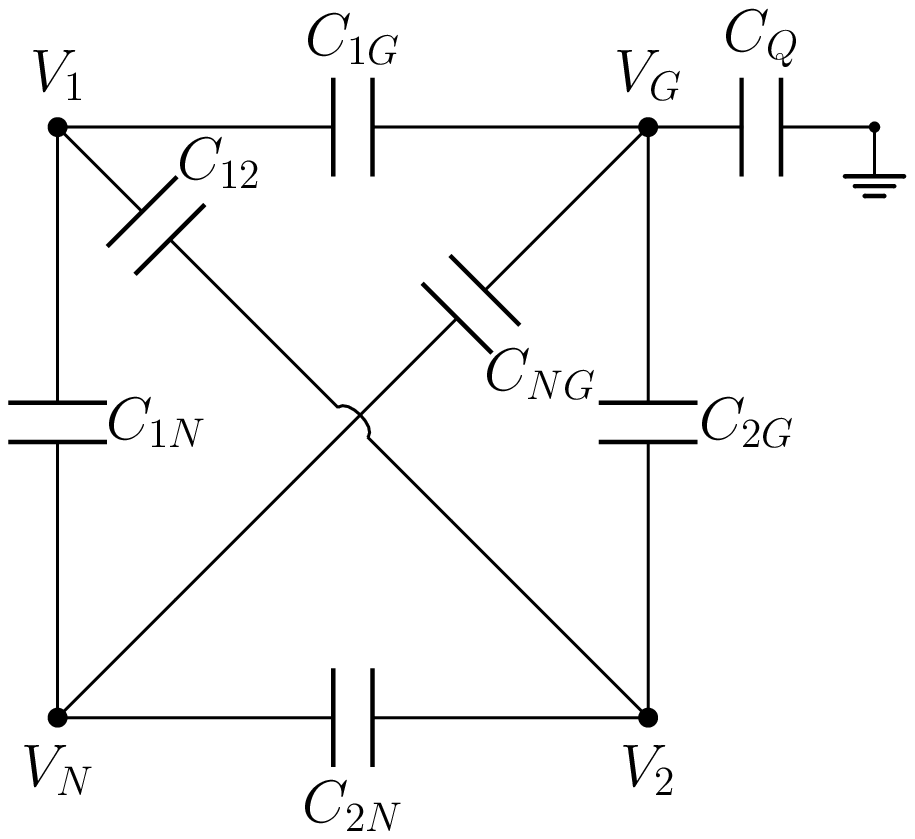}
\label{fig1b}}\caption{(a) Schematic of a graphene sheet subject to $N$
metalic gates. (b) Equivalent circuit plot of (a) with quantum capacitance of
graphene $Q_{C}$ taken into account.}%
\end{figure}

Consider a graphene sheet laid in the $x$-$y$ plain at $z=0$. In the presence
of a dopant concentration $n_{0}(x,y)$ without electric gating, the
quasi-Fermi level is given by%
\begin{equation}
E_{0}(x,y)=\sgn[n_{0}(x,y)]\hbar v_{F}\sqrt{\pi\left\vert n_{0}%
(x,y)\right\vert }, \label{E0}%
\end{equation}
which is obtained from Eq.\ \eqref{n(E)}. When gate voltages of, in general,
$N$ metalic gates are applied as sketched in Fig.\ \ref{fig1a}, the electron
in the graphene layer at $(x,y)$ gains an electrostatic potential energy
$-eV_{G}(x,y)$, where $-e$ is the electron charge and $V_{G}(x,y)=u(x,y,0)$ is
the electrostatic potential $u(x,y,z)$ at $z=0$ to be numerically solved from
the Poisson equation
\begin{equation}
-\nabla\cdot\lbrack\epsilon_{r}(x,y,z)\nabla u(x,y,z)]=\frac{\rho
(x,y,z)}{\epsilon_{0}}, \label{Poisson eq}%
\end{equation}
with $\epsilon_{0}$ the permittivity in free space and $\epsilon_{r}(x,y,z) $
the relative permittivity that can be in principle position dependent.

The energy gain of the electron implies the raising of the energy band of
graphene and, hence, the lowering of the quasi-Fermi level. The graphene
carrier density $n$ therefore obeys Eq.\ \eqref{n(E)} with
\begin{align}
\frac{E(x,y)}{\hbar v_{F}}  &  =\frac{E_{0}(x,y)-[-eV_{G}(x,y)]}{\hbar v_{F}%
}\nonumber\\
&  =\sgn[n_{0}(x,y)]\sqrt{\pi\left\vert n_{0}(x,y)\right\vert }+\frac
{eV_{G}(x,y)}{\hbar v_{F}}, \label{E}%
\end{align}
where $E_{0}(x,y)$ is given by Eq.\ \eqref{E0}. Together with the charges of
the dopant ions that maintain the neutrality of the graphene sheet, the net
charge density on graphene divided by $\epsilon_{0}$ is given by%
\begin{equation}
\left.  \frac{\rho(x,y)}{\epsilon_{0}}\right\vert _{z=0}=\frac{e}{\epsilon
_{0}}\left\{  n_{0}(x,y)-\frac{1}{\pi}\sgn\left[  \bar{E}(x,y)\right]  \bar
{E}(x,y)^{2}\right\}  , \label{BC}%
\end{equation}
where $\bar{E}(x,y)=E(x,y)/\hbar v_{F}$ is given by Eq.\ \eqref{E}. Equation
\eqref{BC} is the boundary condition at the graphene sheet for the Poisson
equation \eqref{Poisson eq}. This boundary condition contains the solution
$V_{G}(x,y)=u(x,y,z=0)$ and, hence, makes the solution process iterative.

\paragraph*{Quantum capacitance model.}

The system of $N$ metalic gates labeled by $j=1,2,\cdots,N$ plus the graphene
sheet labeled by $G$ as sketched in Fig.\ \ref{fig1a} is equivalent to the
circuit plot shown in Fig.\ \ref{fig1b}, where the quantum capacitance of the
graphene sheet $C_{Q}$ is considered. Regarding $G$ as the reference conductor
with electric potential $V_{G}$, the charge density on the surface of each
gate can be expressed as%
\begin{equation}%
\begin{split}
\rho_{1}  &  =C_{1G}(V_{1}-V_{G})+C_{12}(V_{1}-V_{2})+\cdots+C_{1N}%
(V_{1}-V_{N})\\
\rho_{2}  &  =C_{12}(V_{2}-V_{1})+C_{2G}(V_{2}-V_{G})+\cdots+C_{2N}%
(V_{2}-V_{N})\\
&  \vdots\\
\rho_{N}  &  =C_{1N}(V_{N}-V_{1})+C_{2N}(V_{N}-V_{2})+\cdots+C_{NG}%
(V_{N}-V_{G})
\end{split}
, \label{Qgate}%
\end{equation}
where $C_{1G},\cdots,C_{NG}$ are self-partial capacitances and $C_{ij}$ with
$i\neq j$ are mutual partial capacitances.\cite{Cheng1989} Since the whole
isolated system should remain charge neutral, the net charge density on $G$
should be the negative of the total charge density on the $N$ metalic gates:
$\rho_{G}=-\sum_{j=1}^{N}\rho_{j}$. The net electron number density on $G$ is,
therefore, $n_{G}=\rho_{G}/(-e)=\sum_{j=1}^{N}C_{jG}(V_{j}-V_{G})/e$. Suppose
there is an intrinsic doping concentration of $n_{0}$ in graphene. The net
charge density on $G$ is not affected since the number of doped electrons
should equal the number of dopant ions, $\rho_{G}\rightarrow\rho_{G}%
+en_{0}-en_{0}=\rho_{G}$. The net carrier density of graphene, however, is
given by $n=(\rho_{G}-en_{0})/(-e)=n_{G}+n_{0}$, which should obey
Eq.\ \eqref{n(E)}, i.e., $n_{G}+n_{0}=\sgn(E_{0}+eV_{G})[(E_{0}+eV_{G})/\hbar
v_{F}]^{2}/\pi,$ just like in the Poisson-Dirac method. We therefore need to
solve the quadratic equation for $V_{G}$,%
\begin{equation}
\sum_{j=1}^{N}\frac{C_{jG}}{e}(V_{j}-V_{G})+n_{0}=\sgn(E_{0}+eV_{G})\frac
{1}{\pi}\left(  \frac{E_{0}+eV_{G}}{\hbar v_{F}}\right)  ^{2}.
\label{quadraticEQ}%
\end{equation}
After some tedious but straightforward algebra, the carrier density of
graphene in the presence of dopant concentration $n_{0}$ and $N$ gates with
voltages $V_{1},\cdots,V_{N}$ is given by%
\begin{equation}
n=n_{C}+\sgn(n_{C})n_{Q}\left(  1-\sqrt{1+2\frac{\left\vert n_{C}\right\vert
}{n_{Q}}}\right)  +\sgn(n_{0})\sqrt{2n_{Q}\left\vert n_{0}\right\vert },
\label{nQC}%
\end{equation}
where%
\begin{equation}
n_{C}=n_{0}+\sum_{j=1}^{N}\frac{C_{jG}}{e}V_{j} \label{nC}%
\end{equation}
is\ the classical contribution from doping and gating, and
\begin{equation}
n_{Q}=\frac{\pi}{2}\left(  \frac{\hbar v_{F}}{e}\sum_{j=1}^{N}\frac{C_{jG}}%
{e}\right)  ^{2} \label{nQ}%
\end{equation}
arises solely from the quantum capacitance, leading to the second and third
terms in Eq.\ \eqref{nQC} as the quantum correction. Equations
\eqref{nQC}--\eqref{nQ} with $N=1,n_{0}=0,$ and $n_{C}>0$ clearly recover the
results for single-gated pristine graphene given in
Ref.\ \onlinecite{Fang2007}. Contrary to the undoped case,\cite{Fang2007} the
third term in Eq.\ \eqref{nQC} is responsible for the shift of the quasi-Fermi
level due to doping and is typically weak for a reasonable $n_{0}$.

\begin{figure}[b]
\subfigure[]{
\includegraphics[width=\columnwidth]{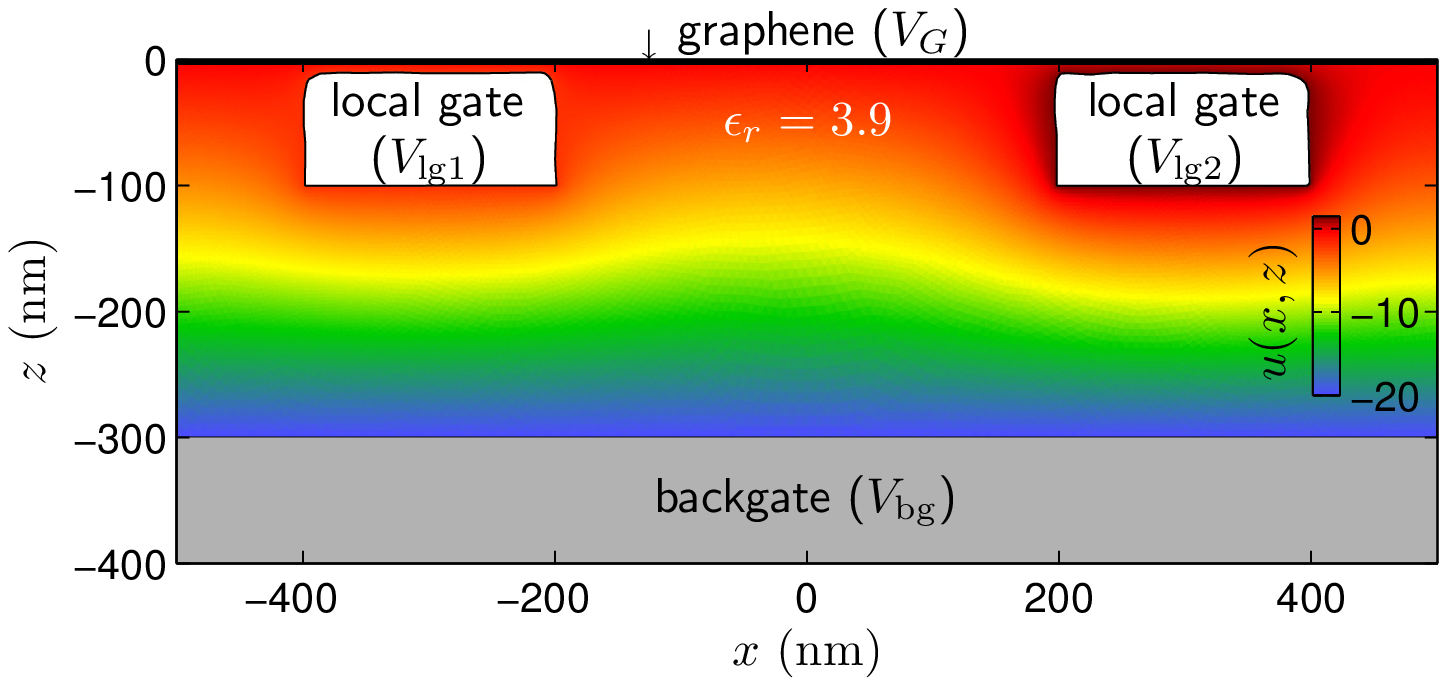}
\label{fig2a}}
\par
\subfigure[]{
\includegraphics[width=\columnwidth]{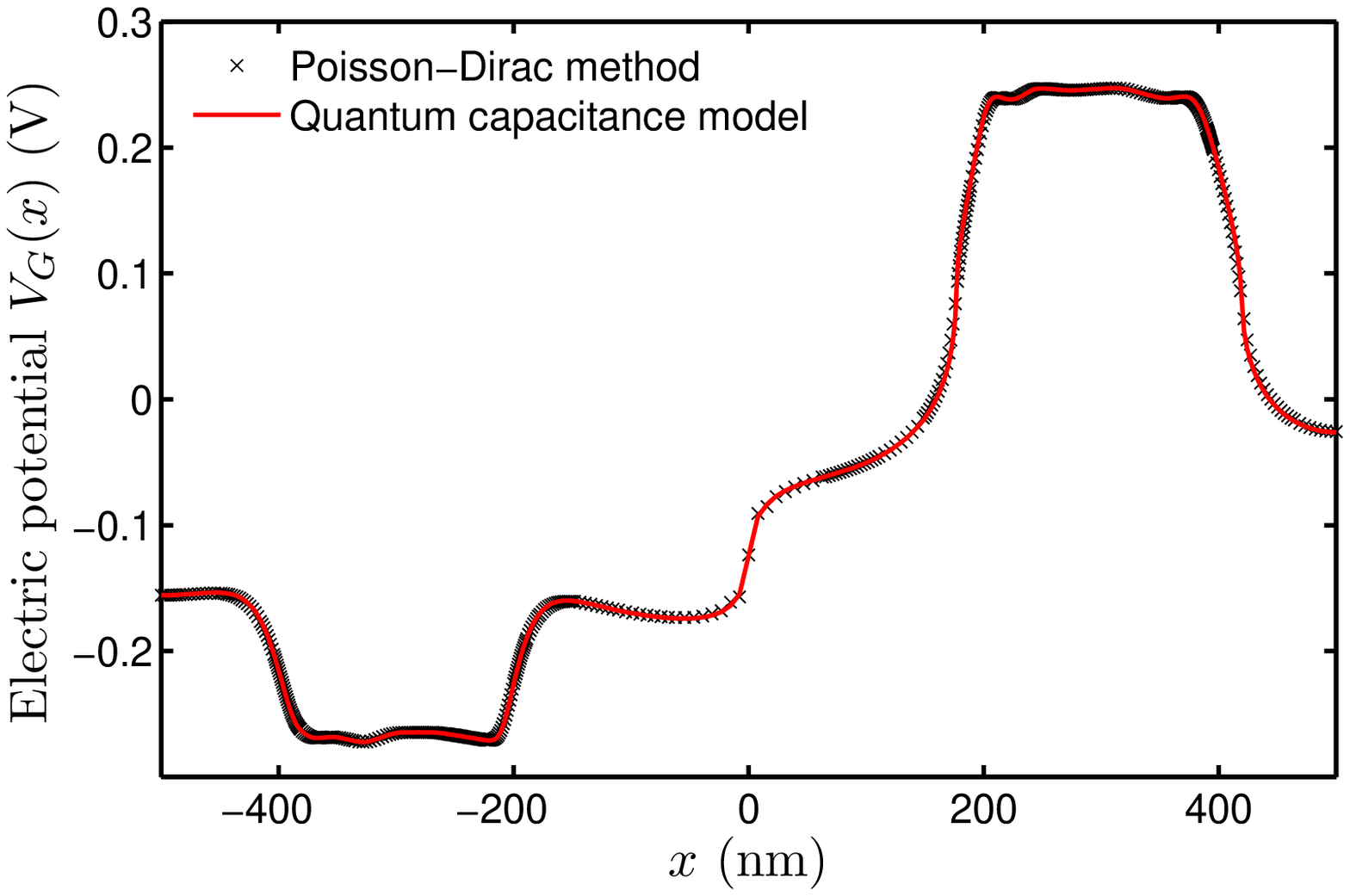}
\label{fig2b}}\caption{(Color online) (a) Side view of a graphene sheet (with
a hyperbolic-tangent-shaped intrinsic doping $n_{0}$) and a backgate (with
$V_{\text{bg}}=-20$ V) sandwiching a SiO$_{2}$ with two embedded local gates
(with $V_{\text{lg1}}=-1.8$ V and $V_{\text{lg2}}=1.5$ V); the color shading
shows the electric potential $u(x,z)$ obtained by the self-consistent
Poisson-Dirac method. (b) The electric potential at the graphene layer
$V_{G}(x)$ obtained by the Poisson-Dirac method and the quantum capacitance
model.}%
\label{fig2}%
\end{figure}

In addition to the doping concentration $n_{0}$ that can have any kind of
spatial profile, the position dependence enters the carrier density
\eqref{nQC} through the self-partial capacitances $C_{1G},C_{2G},\cdots
,C_{NG}$, which can be computed numerically but exactly. For the $i$th gate,
by grounding all the other conductors including the graphene sheet, i.e.,
$V_{j\neq i}=0$ and $V_{G}=0$, Eq.\ \eqref{Qgate} suggests $\bar{n}_{C}%
\equiv-\sum_{j=1}^{N}\rho_{j}/(-e)=(C_{iG}/e)V_{i}$. The self-partial
capacitance for gate $i$ is, therefore, given by%
\begin{equation}
C_{iG}=\left.  \frac{\bar{n}_{C}}{V_{i}}\right\vert _{V_{G}=0,V_{j\neq i}%
=0},\label{CiG}%
\end{equation}
where $\bar{n}_{C}=\pm\epsilon_{r}\epsilon_{0}(\partial u/\partial
z)_{z=0^{\pm}}/e$ can be numerically computed by any kind of finite-element simulator.

With the definitions \eqref{nC} and \eqref{nQ}, one may also write the
solution $V_{G}$ to Eq.\ \eqref{quadraticEQ},%
\begin{equation}
V_{G}=-\frac{\sgn(n_{C})n_{Q}\left(  1-\sqrt{1+2\dfrac{\left\vert
n_{C}\right\vert }{n_{Q}}}\right)  +\sgn(n_{0})\sqrt{2n_{Q}\left\vert
n_{0}\right\vert }}{\sum_{j=1}^{N}\dfrac{C_{jG}}{e}}, \label{VG}%
\end{equation}
which has a reasonable form of charge divided by capacitance, with the
numerator containing only the quantum correction terms in Eq.\ \eqref{nQC}.
The absence of $n_{C}$ in the numerator of $V_{G}$ agrees with our earlier
remark that the classical capacitance model regards graphene as a perfect
conducting plane with fixed zero potential so $n_{C}$ does not contribute to
$V_{G}$.

Equation \eqref{VG} allows for a direct comparison with the iterative solution
obtained from the self-consistent Poisson-Dirac method, as we will show with
an explicit example soon. Multiplying Eq.\ \eqref{VG} with the electron charge
together with the quasi-Fermi level shift $E_{0}$ due to doping,
$-(E_{0}+eV_{G})$ provides for the graphene transport calculation a realistic
on-site energy profile that guarantees a reliable quantum transport
simulation; see, for example, Ref.\ \onlinecite{Liu2012a} for the case with
neglected $n_{0}$. Furthermore, the channel electrostatic potential $V_{G}$
given in Eq.\ \eqref{VG} also allows us to write down the quantum capacitance
of the graphene sheet in the low-temperature limit:\cite{Fang2007}
$C_{Q}\approx(2/\pi)\left(  e/\hbar v_{F}\right)  ^{2}\left\vert
eV_{G}\right\vert $.

\begin{figure}[b]
\centering\includegraphics[width=\columnwidth]{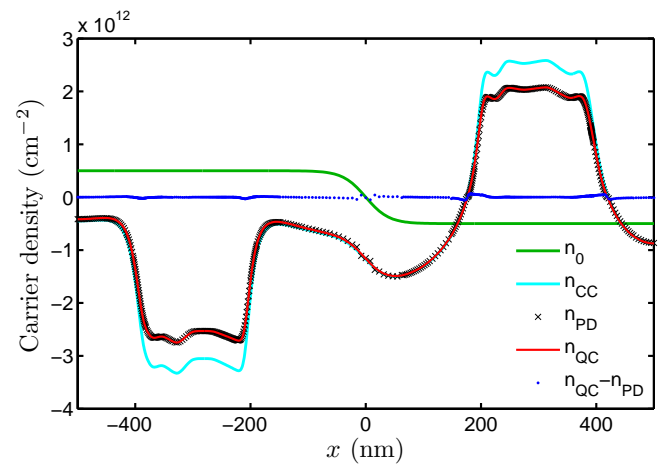}\caption{(Color
online) Carrier density profiles of intrinsic doping $n_{0}(x)$, classical
capacitance model $n_{\text{CC}}(x)$, Poisson-Dirac method $n_{\text{PD}}(x)$,
quantum capacitance model $n_{\text{QC}}(x)$, and the difference
$n_{\text{PD}}(x)-n_{\text{QC}}(x)$, with identical parameters used in
Fig.\ \ref{fig2}.}%
\label{fig3}%
\end{figure}

\paragraph*{Numerical example.}

Armed with the above introduced theories, we next numerically demonstrate the
equivalence of the quantum capacitance model to the self-consistent
Poisson-Dirac iteration method by considering a specific example. To be simple
but general, let us consider a quasi-one-dimensional system along $x$ with
translation invariance along $y$, composed of a doped graphene sheet gated by
one flat backgate and two embedded local gates with irregular shapes roughly
$10$ nm under graphene; see Fig.\ \ref{fig2a}. Embedding such local gates at
such a shallow depth allows independent control of the carrier density in the
locally gated region due to screening of the backgate contribution and can be
experimentally achieved; see, for example, Ref.\ \onlinecite{Nam2011}. The
finite-element method is implemented in the iteration process for the
Poisson-Dirac method as well as the exactly solvable self-partial capacitances
[Eq.\ \eqref{CiG}] for the quantum capacitance model, and the \texttt{pdetool}
in \textsc{Matlab} \cite{pde} is chosen as the simulator for the present demonstration.

The electric potential $u(x,z)$ shown in Fig.\ \ref{fig2a} is obtained by the
self-consistent Poisson-Dirac method with backgate voltage $V_{\text{bg}%
}=-20\mathord{\thinspace\rm V}$ and local gate voltages $V_{\text{lg1}%
}=-1.8\mathord{\thinspace\rm V}$ and $V_{\text{lg2}}%
=1.5\mathord{\thinspace\rm V}$ and an intrinsic doping described by
$n_{0}(x)=-5\times10^{11}\tanh(x/40)\mathord{\thinspace\rm cm}^{-2}$, where
the position coordinate $x$ is in units of nm. The iterated potential solution
$V_{G}(x)=u(x,z=0)$ at the graphene layer is compared in Fig.\ \ref{fig2b}
with the exact solution \eqref{VG} obtained within the quantum capacitance
model, showing an excellent agreement with each other. With other gate
voltages and other shapes of $n_{0}(x)$, the agreement remains exact. Note
that the numerical example chosen here is basically a complicated version of
Ref.\ \onlinecite{Nam2011}, including the proper range of the gate voltages,
except that an artificial doping profile $n_{0}$ with hyperbolic tangent shape
is considered, in order for the comparison to be general.

The spatial profiles of the carrier densities $n_{0}(x)$, $n_{\text{CC}}(x)$,
$n_{\text{PD}}(x)$, $n_{\text{QC}}(x)$, as well as the difference
$n_{\text{PD}}(x)-n_{\text{QC}}(x)$ are shown in Fig.\ \ref{fig3}. Here the
subscripts CC, PD, and QC denote \textquotedblleft classical
capacitance,\textquotedblright\ \textquotedblleft
Poisson-Dirac,\textquotedblright\ and \textquotedblleft quantum
capacitance,\textquotedblright\ respectively. The carrier density within the
classical capacitance model $n_{\text{CC}}$ is obtained by first computing the
induced surface charge at $z=0^{-}$ with the graphene layer grounded
($V_{G}=0$) and then adding the dopant concentration $n_{0}$ or, equivalently,
by Eq.\ \eqref{nC} with the self-partial capacitances [Eq.\ \eqref{CiG}]
numerically computed.

As the quantum correction, i.e., the second and third terms in
Eq.\ \eqref{nQC}, always reduces the magnitude of the net contribution of the
gates, the classical solution always overestimates the gate-induced carrier
density. This correction is especially salient when the gate is close to the
graphene sheet, as is clearly observed by comparing $n_{\text{PD}}(x)$ or
$n_{\text{QC}}(x)$ with $n_{\text{CC}}(x)$ in Fig.\ \ref{fig3}. In addition,
the surface roughness of the embedded local gates considered here with such a
short distance to the graphene sheet (roughly $10\mathord{\thinspace\rm nm}$)
further introduces a strongly fluctuating potential\ profile
[Fig.\ \ref{fig2b}] as well as the corresponding carrier density profile
(Fig.\ \ref{fig3}) at the locally gated regions.

As in the case of $V_{G}(x)$ compared in Fig.\ \ref{fig2b}, the agreement
between $n_{\text{PD}}(x)$ and $n_{\text{QC}}(x)$ is rather satisfactory. In
Fig.\ \ref{fig3}, the discrepancy between the Poisson-Dirac method and the
quantum capacitance model becomes relatively obvious near positions where the
surface charge density of the boundary condition \eqref{BC} is changing its
sign. This implies that the discrepancy may stem from the inherent numerical
limitation of the chosen nonlinear partial differential equation solver.

\paragraph*{Conclusion.}

In conclusion, an exact solution for the space-resolved carrier density in
multigated doped graphene sheets within the quantum capacitance model has been
derived. With an illustrative quasi-one-dimensional example, the exact
solution is shown to be equivalent to the self-consistent Poisson-Dirac
iteration method. The solution therefore provides a fast and accurate way to
compute spatially varying carrier density, on-site potential energy (key input
for quantum transport simulation), as well as quantum capacitance for bulk
graphene, allowing for any kind of gating geometry and any types of intrinsic
doping. Moreover, the contact doping\cite{Giovannetti2008,Khomyakov2009} and
its corresponding screening potential\cite{Khomyakov2010} can as well be
treated by the presented solution, which therefore takes care of all three
types of doping in graphene---electric, chemical, and contact-induced---in a
unified manner.

\paragraph*{Acknowledgments.}

The author thanks K.\ Richter for valuable discussions and suggestions.
Financial support by the Alexander von Humboldt foundation is gratefully acknowledged.

\end{CJK*}

\thispagestyle{plain}

\bibliographystyle{apsrev4-1}
\bibliography{mhl2}
\end{document}